\documentclass[twocolumn,preprintnumbers,superscriptaddress,prl,nofootinbib]{revtex4-1}
\usepackage{graphicx}
\usepackage{times}
\usepackage{epsfig}
\usepackage{epstopdf}
\usepackage{amsmath}
\usepackage{amsfonts}
\usepackage{amssymb}
\usepackage{url}
\usepackage{subfigure}
\usepackage{verbatim}
\usepackage{comment}

\newcommand{\be}{\begin{equation}}
\newcommand{\ee}{\end{equation}}
\newcommand{\bea}{\begin{eqnarray}}
\newcommand{\eea}{\end{eqnarray}}

\begin{document}
%\preprint{}
% Page numbers bottom-center
\pagestyle{plain}
%%%%%%%%%%%%%%%%%%%%%%%%%%%%%%%%%%%%%%%%%%%%%%%%%%%%%%%%%%%%
\title{
Dark Matter $Z^\prime$ and XENON1T Excess from $U(1)_X$ Extended Standard Model 
}
%%%%%%%%%%%%%%%%%%%%%%%%%%%%%%%%%%%%%%%%%%%%%%%%%%%%%%%%%%%%%
%\author{Nobuchika Okada}
%\email{okadan@ua.edu}
%\affiliation{
%Department of Physics and Astronomy,
%University of Alabama,
%Tuscaloosa, AL 35487, USA
%}
%\author{Qaisar Shafi}
%\Email{shafi@bartol.udel.edu}
%\affiliation{
%Bartol Research Institute,
%Department of Physics and Astronomy,
%University of Delaware, Newark, DE 19716, USA
%}
%
\author{Nobuchika Okada}
\email{okadan@ua.edu}
\affiliation{
Department of Physics and Astronomy,
University of Alabama,
Tuscaloosa, AL 35487, USA
}
\author{Satomi Okada}
\email{satomi.okada@ua.edu}
\affiliation{
Department of Physics and Astronomy,
University of Alabama,
Tuscaloosa, AL 35487, USA
}
\author{Digesh Raut}
\email{draut@udel.edu}
\affiliation{
Bartol Research Institute,
Department of Physics and Astronomy,
University of Delaware, Newark, DE 19716, USA
}
\author{Qaisar Shafi}
\email{qshafi@udel.edu}
\affiliation{
Bartol Research Institute,
Department of Physics and Astronomy,
University of Delaware, Newark, DE 19716, USA
}

%\date{\today}

%\baselineskip 36pt

%%%%%%%%%%%%%%%%%%%%%%%%%%%%%%%%%%%%%%%%%%%%%%%%%%%%%%%%%%
\begin{abstract}

A gauged $U(1)_X$ symmetry appended to the Standard Model (SM) is particularly well motivated since it can account for the light neutrino masses by the seesaw mechanism, 
explain the origin of baryon asymmetry of the universe via leptogenesis, 
  and help implement successful cosmological inflation with the $U(1)_X$ breaking Higgs field as the inflaton. 
In this framework, we propose a light dark matter (DM) scenario in which the $U(1)_X$ gauge boson $Z^\prime$ behaves as a DM particle in the universe. 
We discuss how this scenario with $Z^\prime$ mass of a few keV
  and a $U(1)_X$ gauge coupling $g_X \simeq 10^{-16}$ can nicely fit the excess 
  in the electronic recoil energy spectrum recently reported by the XENON1T collaboration. 
In order to reproduce the observed DM relic density in the presence of such a tiny gauge coupling, 
we propose an extension of the model to a two-component DM scenario.  
The $Z^\prime$ DM density can be comparable to the observed DM density 
  by the freeze-in mechanism through the coupling of $Z^\prime$ boson to a partner Higgs-portal scalar DM with a large $U(1)_X$ charge. 
 
\end{abstract}
\maketitle

%%%%%%%%%%%%%%%%%%%%%%%%

Over the years the direct dark matter (DM) detection experiments have continuously improved 
  with increased detector volume and sensitivity. 
The primary target for the search is the so-called Weakly Interacting Massive Particle (WIMP) DM
  through its elastic scattering off nucleons. 
Since no evidence for DM has been found, the cross section of the DM particle  
  with mass $\gtrsim 1$ GeV is very severely constrained. 
For example, the XENON1T experiment \cite{Aprile:2018dbl} has set an upper bound on the spin-independent 
  elastic scattering cross section with a nucleon as $\sigma_{SI} \lesssim 4 \times 10^{-11}$ pb 
  for a 30 GeV WIMP DM. 
Due to its unprecedentedly low background rate, large target mass and low energy threshold, 
  the XENON1T experiment can also explore alternative dark matter candidates 
  with mass in the range of 1-100 keV by using the electronic-recoil events. 
It has recently reported an excess of the electronic-recoil events 
  below $7$ keV, which is prominent around a few keV \cite{Aprile:2020tmw} with a local statistical significance of 3-4 $\sigma$.

The XENON1T collaboration has examined a possible explanation of the excess 
   by solar axions, an enhanced neutrino magnetic moment using solar neutrinos, and bosonic (pseudo-scalar and vector) DM particles.   
The identified parameter regions for the solar axion and the neutrino magnetic moment signals 
   are both in strong tension with astrophysical constraints, in particular the stellar cooling constraints 
   (see, for example, Ref.~\cite{DiLuzio:2020jjp} and references therein). 
The excess can also be explained by $\beta$ decay of tritium, which was initially not considered, 
   if a sufficient amount of tritium survived the xenon purification process \cite{Aprile:2020tmw, Robinson:2020gfu}. 
If this is the case, the significance of the observed excess is reduced to around 2 $\sigma$, 
   and this analysis sets the most restrictive direct constraints to date 
   on pseudo-scalar and vector DM particles with masses between $1$ and $210$ keV.
Other possible backgrounds are discussed in Ref.~\cite{Bhattacherjee:2020qmv}.

Following the report by the XENON1T collaboration, many works have appeared  
   which attempt to explain the XENON1T excess based on new physics scenarios 
   with Axion Like Particles (ALPs) \cite{Takahashi:2020bpq}, DM particles \cite{Kannike:2020agf, Alonso-Alvarez:2020cdv, An:2020bxd}, 
   and others \cite{Amaral:2020tga}. 
In this letter we propose a light DM scenario in a gauged $U(1)_X$ extension of the Standard Model (SM), 
   in which the $U(1)_X$ gauge boson $Z^\prime$ serves as the DM in the universe. 
With a suitable choice of its mass and $U(1)_X$ gauge coupling values, 
   the $Z^\prime$ DM can account for the XENON1T excess. 
This $U(1)_X$ model can also provide us with a natural explanation of tiny neutrino masses by the seesaw mechanism \cite{Seesaw}
   and the origin of the observed matter-anitmatter asymmetry in the universe via leptogenesis \cite{Fukugita:1986hr}. 
Furthermore, a successful cosmological inflation scenario can be realized with non-minimal gravitational coupling \cite{Okada:2014lxa}
 after identifying the $U(1)_X$ Higgs field with the inflaton \cite{Okada:2011en}.

It has been shown in Refs.~\cite{Alonso-Alvarez:2020cdv, An:2020bxd} 
   that if it is the dominant DM component in our universe, 
   a dark photon with mass ($m_{A^\prime}$) around a few keV and a kinetic mixing $\epsilon \sim10^{-15}$ with the SM photon  
   can nicely fit the observed recoiled-electron energy spectrum observed by the XENON1T experiment, 
   while satisfying the astrophysical constraints.   
Here we refer to the best fit value in Ref.~\cite{Alonso-Alvarez:2020cdv}: 
   $m_{A^\prime} = 2.8$ keV and the kinetic mixing $\epsilon = 8.6 \times 10^{-16}$.  
This parameter set can not only fit the XENON1T excess but also account for the observed anomalous cooling of horizontal branch stars
   by resonant production of dark photons in the stellar interior.
As we will show in the following, the coupling of $Z^\prime$ DM with (right-handed) electron 
   is essentially the same as that of the dark photon induced by its kinetic mixing with photon, 
   so that we have a relation, $g_X=e \epsilon$, 
   where $g_X$ and $e$ are the $U(1)_X$ gauge and electromagnetic couplings, respectively. 
Therefore, we can interpret the dark photon result to our $Z^\prime$ DM case 
   and conclude that our model can fit the XENON1T excess 
   by setting the $Z^\prime$ DM mass $m_{Z^\prime}=2.8$ keV 
   and the $U(1)_X$ gauge coupling $g_X =1.8 \times 10^{-16}$.

There is a common issue among the DM models which have been proposed to explain the XENON1T excess: 
   how can we reproduce the observed DM relic density of $\Omega_{DM} h^2=0.12$ \cite{Planck2018}?  
With model parameters set to account for the XENON1T excess, 
   we may evaluate a thermal or non-thermal DM density by the freeze-out or freeze-in mechanism. 
We find that the resultant DM density is most likely too large in thermal DM models 
 and too small in non-thermal DM models.  
In the minimal $U(1)_X$ extension of the SM, we also encounter this issue 
   since the $Z^\prime$ DM coupling with the SM particles is extremely weak
   to account for the XENON1T excess. 
We propose a modest extension of the minimal model to a two-component DM scenario, 
   in which the $Z^\prime$ DM density can be compatible to the observed DM density 
   by the freeze-in mechanism. 
This is accomplished through a $Z^\prime$ DM coupling to its partner Higgs-portal scalar DM
   with a large $U(1)_X$ charge.

The $Z^\prime$ DM interpretation to the XENON1T excess 
   with $m_{Z^\prime} = {\cal O}$(1 keV) and $g_X ={\cal O}(10^{-16})$ 
   indicates the $U(1)_X$ spontaneous symmetry breaking scale to be 
   $v_X \simeq m_{Z^\prime}/g_X =  {\cal O}(10^{10} \, {\rm GeV})$. 
This is very interesting in view of the fact 
   that this $U(1)_X$ symmetry breaking scale is same order as the mass scale 
   of right-handed neutrinos (RHNs) in the $U(1)_X$ extended SM. 
As discussed in Ref.~\cite{LGbound}, 
   successful baryogenesis via the standard thermal leptogenesis   
   requires the lightest RHN mass to be $M_N \gtrsim 10^9$ GeV. 
This constraint is satisfied by the symmetry breaking scale derived from the parameter set 
   accounting for the XENON1T excess. 
Of course, successful thermal leptogenesis requires the reheating temperature 
   after inflation $T_{RH} >  M_N$. 
In the following, we will also briefly discuss a simple inflation scenario with the $U(1)_X$ Higgs field as inflaton 
   and show that the inflaton decay can reheat the universe to the desired temperature.

%%%%%%%%%%%%%%%%%%%%%%%%
\begin{table}[t]
\begin{center}
\begin{tabular}{|c|ccc|c|}
\hline
      &  $SU(3)_c$  & $SU(2)_L$ & $U(1)_Y$ & $U(1)_X$  \\ 
\hline
$q^{i}_{L}$ & {\bf 3 }    &  {\bf 2}         & $ 1/6$       & $(1/6) x_{H} + (1/3)$   \\
$u^{i}_{R}$ & {\bf 3 }    &  {\bf 1}         & $ 2/3$       & $(2/3) x_{H} + (1/3)$   \\
$d^{i}_{R}$ & {\bf 3 }    &  {\bf 1}         & $-1/3$       & $(-1/3) x_{H} + (1/3)$  \\
\hline
$\ell^{i}_{L}$ & {\bf 1 }    &  {\bf 2}         & $-1/2$       & $(-1/2) x_{H} -1$    \\
$e^{i}_{R}$    & {\bf 1 }    &  {\bf 1}         & $-1$                   & $- x_{H} -1$   \\
\hline
$H$            & {\bf 1 }    &  {\bf 2}         & $- 1/2$       & $(-1/2) x_{H}$   \\  
\hline
$N^{i}_{R}$    & {\bf 1 }    &  {\bf 1}         &$0$                    & $-1$     \\
$\Phi$            & {\bf 1 }       &  {\bf 1}       &$ 0$                  & $ 2$  \\ 
\hline
\end{tabular}
\end{center}
\caption{
The particle content of the minimal U(1)$_X$ extended SM, 
  where $i = 1,2,3$ is the generation index, and $x_H$ is a real parameter. 
The model is free from all the gauge and mixed gauge-gravitational anomalies. 
 }
\label{tab:1}
\end{table}
%%%%%%%%%%%%%%%%%%%%%%%%%%

Let us now present more details of our model. 
It is based on the minimal $U(1)_X$ extended SM \cite{Appelquist:2002mw}, 
  which is a generalization of the minimal $B-L$ (baryon number minus lepton number) model \cite{mBL}
  in which the anomaly-free global $B-L$ symmetry of the SM 
  is gauged, and three RHNs and a $U(1)_{B-L}$ charged Higgs field are introduced. 
The model is free from all the gauge and mixed gauge-gravitational anomalies 
  in the presence of the three RHNs.  
In the $U(1)_X$ generalization, the $U(1)_X$ charge of a field is determined by 
  a linear combination of its SM hyper-charge and $B-L$ charge: 
  $Q_X= x_H \, Q_Y+ Q_{B-L}$, where $x_H$ is a real parameter. 
Since both the SM $U(1)_Y$ and the $U(1)_{B-L}$ symmetries are anomaly free
  in the minimal $B-L$ model, the $U(1)_X$ gauge symmetry is anomaly-free
  with the charge assignment as the linear combination.  
The particle content of the minimal $U(1)_X$ model is summarized in Table \ref{tab:1}.

In the model we introduce the Higgs potential of the form: 
\bea
 V &=& \lambda_X \left(\Phi^\dagger \Phi - \frac{v_X^2}{2} \right)^2 
   +  \lambda_H \left(H^\dagger H - \frac{v_h^2}{2} \right)^2    \nonumber \\
   &+& \lambda_{mix} \left(\Phi^\dagger \Phi - \frac{v_X^2}{2} \right) \left(H^\dagger H - \frac{v_h^2}{2} \right). 
\label{potential}   
\eea
The ground state of the system is determined by the vacuum expectation values (VEVs) of 
   $\langle \Phi \rangle = v_X/\sqrt{2}$ and $\langle H \rangle = (v_h/\sqrt{2}, 0)^T$, 
   where $v_h=246$ GeV is the SM Higgs VEV.  
We assume $\lambda_{mix}$ is small and the mixing between the $U(1)_X$ Higgs and SM Higgs fields is negligible. 
As previously mentioned, $v_X \sim 10^{10}$ GeV$\gg v_h$, so that the $Z^\prime$ boson mass 
   and the $U(1)_X$ Higgs boson mass are given by 
\bea
m_{Z^\prime}  =  2 g_{X} v_{X},  \qquad m_\phi = \sqrt{2 \lambda_X } v_X,  
\eea
respectively. 
In addition to the Yukawa couplings in the SM model, we have the new Yukawa couplings: 
\bea
\mathcal{L}_{Y} \supset  - \sum_{i, j=1}^{3} Y^{ij}_{D} \overline{\ell^i_{L}} H N_R^j 
          -\frac{1}{2} \sum_{k=1}^{3} Y_M^k \Phi \overline{N_R^{k~C}} N_R^k,  
       %+ {\rm h.c.} ,
\label{eq:Lseesaw} 
\eea
where $Y_D$ and $Y_M$ are Dirac and Majorana-type Yukawa coupling matrices, respectively, 
  and we have chosen flavor-diagonal Majorana Yukawa couplings without loss of generality. 
Associated with the electroweak and $U(1)_X$ gauge symmetry breaking, 
  the Dirac and Majorana mass matrices are generated as 
  $m_D=Y_D v_h/\sqrt{2}$ and $M_N=Y_M v_X/\sqrt{2}$.   
Assuming a hierarchy between the scales of the two mass matrix elements, 
  the light neutrino mass matrix $M_\nu \simeq m_D M_N^{-1} m_D^T$ 
  is generated by the seesaw mechanism and the tiny neutrino mass scale is naturally derived.   
For $Y_M^k = {\cal O}(1)$, the seesaw scale is set by $v_X$.

We now propose the light $Z^\prime$ DM scenario in the minimal $U(1)_X$ extended SM by setting $x_H=-2$. 
From Table \ref{tab:1} we can see that with this choice the left-handed SM fermions
  are neutral under the $U(1)_X$ symmetry. 
This means that if $m_{Z^\prime} < 2 \, m_e$, where $m_e =0.51$ MeV is the electron mass, 
  the $Z^\prime$ boson becomes a long-lived DM candidate 
  since it cannot decay into neutrinos (in the flavor basis).\footnote{Here we assume that the tree-level kinetic mixing between the $U(1)_X$ and $U(1)_Y$ gauge fields is zero. Although quantum corrections can generate such a kinetic mixing, the corresponding mixing parameter is estimated to be of order $\sim \frac{1}{16 \pi^2}\frac{g_X^2}{ g_Y^2}$, which is negligibly small. Therefore, we can safely ignore its effect in our analysis.} 
The right-handed electron has the gauge coupling with the $Z^\prime$ DM: 
\bea
{\cal L}_{int} = g_X  \left( {\overline e_R} \gamma^\mu e_R \right)  Z^\prime_\mu. 
\eea
Note that this is essentially the same as the dark photon coupling with the electron, 
\bea
 {\cal L}_{int} = \epsilon \, e \,  \left( {\overline e} \gamma^\mu e \right)  A^\prime_\mu, 
\eea
except that the left-handed electron has no $Z^\prime$ gauge interaction. 
Therefore, we can interpret the best fit values for the dark photon parameters
  to account for the XENON1T excess \cite{Alonso-Alvarez:2020cdv} as 
\bea
   m_{Z^\prime}=m_{A^\prime} =2.8 \, {\rm keV} , \qquad 2 g_X^2 = \epsilon^2 e^2.  
\eea
Here, since the left-handed electron is not involved in the $Z^\prime$ DM absorption process,
  we have added the factor 2 in the coupling relation 
  to yield the same absorption rate as the dark photon case. 
Using the fine-structure constant $\alpha_{em}=\frac{e^2}{4 \pi}=\frac{1}{137}$ at low energies,
  the best fit value for the $U(1)_X$ gauge coupling is found to be $g_X=1.8 \times 10^{-16}$.

Similar to the dark photon, the $Z^\prime$ DM is not stable. 
The main decay process of the dark photon is $A^\prime \to \gamma \gamma \gamma$
  through quantum corrections with electron loops. 
The partial decay width is calculated in Ref.~\cite{Redondo:2008ec}, 
   which is interpreted in the $Z^\prime$ DM case as 
\bea
\Gamma_{Z^{\prime}\to 3 \gamma} \simeq 3.1 \times 10^{-30} {\rm Gyr}^{-1} 
 \left(\frac{m_Z^\prime} {2.8 \;{\rm keV}}\right)^9 \left(\frac{g_X}{10^{-16}}\right)^2. 
\eea
In the $Z^\prime$ DM case, a new decay mode arises from its gauge coupling with RHNs.  
%\bea
%{\cal L}_{int} \supset - g_X \left( \sum_j \, {\overline N_R^j} \gamma^\mu N_R^j  \right) Z^\prime_\mu. 
%\eea
In the flavor basis, the left-handed neutrinos have no $U(1)_X$ gauge interaction, 
  but the light neutrino mass eigenstates have  couplings with $Z^\prime$ DM 
  through the mass mixings generated by the seesaw mechanism. 
The mixing is roughly estimated to be $m_D/M_N$ with $m_D$  and $M_N$ 
  being the typical mass scales of the Dirac and Majorana mass matrix elements. 
The $Z^\prime$ DM coupling with a light neutrino mass eigenstate is then expressed as 
\bea
{\cal L}_{int} \sim  g_X   \left( \frac{m_D}{M_N} \right)^2 
 \left({\overline \nu}  \gamma^\mu  \gamma_5 \nu \right)  {Z^\prime}_\mu,  
\label{eq:LZNN}
\eea
where $\nu$ is the four-component spinor representation of the light neutrino  Majorana mass eigenstate. 
From the seesaw formula, we expect $(m_D/M_N)^2 \sim \sqrt{\Delta m_{23}^2}/M_N$,  
   where $\Delta m_{23}^2 \simeq 2.4 \times 10^{-3} \, {\rm eV}^2$ 
   is the neutrino oscillation data \cite{Tanabashi:2018oca}. 
We now evaluate the partial $Z^\prime\to \nu \nu$ decay width as 
\bea
\frac{\Gamma_{Z^{\prime}\to \nu \nu}}{\Gamma_{Z^{\prime}\to 3 \gamma}} 
 \simeq  \left(\frac{m_Z^\prime} {2.8 \; {\rm keV}}\right)^8
\left(\frac{4.6 \times 10^5 \;{\rm GeV}}{M_N}\right)^2.  
\eea
For $M_N \lesssim 4.6 \times 10^5$ GeV, the $Z^\prime$ DM mainly decays to a pair of neutrinos. 
In any case, the $Z^\prime$ DM is sufficiently long-lived.

Next we evaluate the $Z^\prime$ DM density at present. 
Since its coupling with the SM particles is too weak for the $Z^\prime$ DM to be in thermal equilibrium in the early universe, 
  we consider the $Z^\prime$ DM production by the freeze-in mechanism
  through $\overline{f_{SM}} \, f_{SM} \to Z^\prime \, \gamma$, where $f_{SM}$ denotes a SM fermion.
To calculate the relic density of $Z^\prime$ DM, we solve the Boltzmann equation 
\bea
\frac{dY_{Z^\prime}}{dx}\simeq \frac{ \langle \sigma v \rangle}{x^2}\frac{s( m_{Z^\prime})}{H( m_{Z^\prime})} Y^{eq}_{Z^\prime} Y^{eq}_{\gamma}. 
\label{eq:Boltzmann}
\eea
Here $x \equiv m_{Z^\prime}/T$, and the entropy density ($s$), 
  the Hubble parameter ($H$) and the equilibrium yields of $Z^\prime$ ($Y^{eq}_{Z^\prime}$) and $\gamma$ ($Y^{eq}_{\gamma}$) 
  are given by
\bea
&&H(m_{Z^\prime}) = \sqrt {\frac{\pi^2}{90}g_*} \frac{m_{Z^\prime}^2}{M_P},  \qquad 
s(m_{Z^\prime}) = \frac{2\pi^2}{45}g_* m_{Z^\prime}^3, 
\nonumber \\
&&Y^{eq}_{Z^\prime} \sim Y^{eq}_{\gamma} \simeq 2  \frac{45}{2\pi^4} \frac{1}{g_*} \simeq 4.4 \times 10^{-3},  \;\;\; {\rm for}\; x \lesssim 1, 
\eea
where $M_P=2.43 \times 10^{18}$ GeV is the reduced Planck mass. 
We have set the effective relativistic degrees of freedom $g_* \simeq 106.75$. 
The annihilation/creation cross section for the process ${\overline {f_{SM}} \, f_{SM} \leftrightarrow Z^\prime \, \gamma}$ 
   is roughly given by
\bea
\langle \sigma v \rangle \simeq \frac{g_X^2 \alpha_{EM}}{m_{Z^\prime}^2} x^2, \qquad {\rm for}\;x \lesssim 1.
\eea
Using this we can easily find the solution, 
\bea
Y_{Z^\prime} (x) \simeq 2.6 \times 10^{-4} \left(\frac{M_P}{m_{Z^\prime}}\right) g_X^2 \alpha_{em} (x - x_{RH}), 
\eea
where $x_{RH}=m_{Z^\prime}/T_{RH} \ll 1$, $T_{RH}$ is the reheating temperature  
  after inflation, and we have assumed $Y(x_{RH})=0$.  
The $Z^\prime$ production effectively stops at $T \simeq m_e$, 
  since for $T <m_e$ the electron number density is suppressed by the Boltzmann factor. 
Thus, we estimate the DM yield at present by $Y(\infty) \simeq Y(x_e=m_{Z^\prime}/m_e)$. 
The resultant yield is independent of $x_{RH} \ll m_{Z^\prime}/m_e$.  
For the freeze-in mechanism, this rough estimate is found to well-approximate the numerical solution
  (see, for example, Ref.~\cite{Okada:2020cue}).  
The $Z^\prime$ DM relic density at present is given by
\bea
\Omega_{Z^\prime} h^2 &=& \frac{m_{Z^\prime} s_0 Y_{Z^\prime} ( \infty)}{\rho_c/h^2}
 \simeq \frac{m_{Z^\prime} s_0 Y_{Z^\prime} (x_e)}{\rho_c/h^2} \nonumber\\
 &\simeq& 6.6 \times 10^{-14}
 \left(\frac{g_X}{10^{-16}}\right)^2 
\left(\frac{m_{Z^\prime}}{2.8 \; {\rm keV}}\right),  
\label{eq:relic}
\eea
where $\rho_c/h^2 =1.05 \times 10^{-5}$ and $s_0 = 2890/{\rm cm}^3$. 
This clearly is an extremely tiny relic density. 
To achieve the observed relic density $\Omega_{DM} h^2 =0.12$, 
  we may consider some non-thermal $Z^\prime$ production mechanism. 
See, for example, Ref.~\cite{Alonso-Alvarez:2020cdv} and references therein.

In this letter we propose another possibility to alleviate this relic density problem. 
We extend the minimal model by introducing a SM singlet scalar field ($\varphi$) with a $U(1)_X$ charge $Q_\varphi$. 
If $Q_\varphi \neq \pm 2$, this scalar is stable and hence another DM candidate in our model, which can be identified with the well-known Higgs-portal DM, 
  in which the scalar DM communicates with the SM sector through its unique renormalizable coupling 
  with the SM Higgs doublet, ${\cal L}_{int} =\lambda_\varphi (\varphi^\dagger \varphi)(H^\dagger H)$. 
Although this scenario is very severely constrained by the direct and indirect DM detection experiments, 
  two typical mass regions, $m_\varphi \simeq m_H/2 \simeq 62.5$ GeV and $m_\varphi \gtrsim 2$ TeV are compatible with $\Omega_\varphi h^2 \leq 0.12$.\footnote{For the coupling $\lambda_\varphi$ within the perturbative regime, 
the currently allowed parameter space will be probed by the next generation direct detection experiments such as LUX-ZEPLIN (LZ), XENONnT and DARWIN. For a recent review, see Ref.~\cite{Arcadi:2019lka}.}
Since $Q_\varphi$ is a model parameter, we may choose $|Q_\varphi| \gg 1$. 
In this case, the gauge interaction between the Higgs-portal DM $\varphi$ and the $Z^\prime$ DM is increased, 
  $g_\varphi \equiv |Q_\varphi| g_X \gg g_X\simeq 10^{-16}$, 
  and the $\varphi$ DM assists to achieve $\Omega_{Z^\prime} h^2 =0.12$ in the freeze-out mechanism.

Through the Higgs-portal coupling, the $\varphi$ DM is in thermal equilibrium with the SM particles 
  until it decouples at a typical decoupling temperature for a WIMP DM, $T_{dec} \sim  m_\varphi/20$.    
In the early universe, the $Z^\prime$ DM particles  are produced by pair annihilation of the Higgs-portal DM particles
  through the $U(1)_X$ gauge interaction, $\varphi^\dagger \varphi \leftrightarrow Z^\prime Z^ \prime$. 
We estimate the annihilation/creation cross section to be 
\bea
\langle \sigma v \rangle \simeq \frac{g_\varphi^4}{4 \pi}\frac{x^2}{m_{Z^\prime}^2},  
  \qquad {\rm for}\;  x \lesssim  \frac{m_{Z^\prime}}{m_\varphi},    
\label{eq:DMcs2}
\eea
before the decoupling of $\varphi$ DM from the SM thermal plasma. 
To estimate the $Z^\prime$ DM density, we solve the Boltzmann equation with this cross section: 
\bea
\frac{dY_{Z^\prime}}{dx}\simeq \frac{ \langle \sigma v \rangle}{x^2}\frac{s( m_{Z^\prime})}{H( m_{Z^\prime})} \left({Y^{eq}_{Z^\prime}}\right)^2, 
\label{eq:Boltzmann2}
\eea
which can be easily solved. 
Similar to the previous analysis, the $Z^\prime$ DM production effectively stops at $T \simeq m_\varphi$ 
  since the number density of $\varphi$ is Boltzmann suppressed for $T < m_\varphi$.  
We then estimate the present yield to be
\bea
Y_{Z^\prime}(\infty) \simeq Y_{Z^\prime} \left(x_\varphi \right) 
\simeq
2.1 \times 10^{-5} \left(\frac{M_P}{m_\varphi}\right) g_\varphi^4,  
\eea
where $x_\varphi \equiv m_{Z^\prime}/m_\varphi$, 
and the resultant relic density is 
\bea  
\Omega_{Z^\prime} h^2 \simeq 6.4 \times 10^{14} \, g_\varphi^4
\left(\frac{m_{Z^\prime}}{2.8 \; {\rm keV}}\right)
\left(\frac{62.5 \; {\rm GeV}}{m_\varphi}\right). 
\eea 
For $m_\varphi=m_H/2 = 62.5$ GeV and $m_{Z^\prime}=2.8$ keV, 
  we find $g_\varphi = 1.2 \times 10^{-4}$ to reproduce $\Omega_{Z^\prime} h^2 =0.12$. 
Although this $g_\varphi$ value corresponds to a huge $U(1)_X$ charge $|Q_\phi| \sim 10^{12}$ for $g_X \sim 10^{-16}$, 
  there is no apparent problem with particle physics phenomenology.

Since we have two DM candidates, $Z^\prime$ and $\varphi$, 
  it is not necessary for the $Z^\prime$ DM to account for the total $\Omega_{DM}h^2=0.12$. 
We define $Z^\prime$ DM fraction to the total DM density as 
  $R= \Omega_{Z^\prime} h^2/0.12$. 
Considering that the observed XENON1T signal events are proportional to $g_X^2 R$, 
  the best fit value of the $U(1)_X$ gauge coupling scales with $R$: 
\bea
    g_X = \frac{1.8 \times 10^{-16}} {\sqrt{R}}. 
\eea    
According to the stellar cooling constraints shown in Refs.~\cite{Alonso-Alvarez:2020cdv, An:2020bxd}, 
  we find a lower bound on the fraction, $R \gtrsim 0.01$.

Our model can also account for the origin of the baryon asymmetry of the universe 
  via leptogenesis through the out-of-equilibrium decay of the RHNs. 
The successful standard thermal leptogenesis requires the lightest RHN mass $M_N \gtrsim 10^{9}$ GeV 
  and the reheating temperature $T_{RH} \gtrsim M_N$. 
The best fit values for the XENON1T excess, $m_{Z^\prime}=2.8$ keV and $g_X=1.8 \times 10^{-16}$
  lead to $v_X=7.8 \times 10^9$ GeV, and 
  the RHN mass bound is satisfied if we take $Y_M \gtrsim 0.1$.  
The reheating temperature depends on inflation models. 
In the following, we briefly discuss how our $U(1)_X$ model provides a successful inflation scenario 
  with the $U(1)_X$ Higgs field as the inflaton.  
The latter decays into the SM Higgs doublet which reheats the universe to the desired temperature for leptogenesis.

We consider quartic inflation with non-minimal gravitational coupling 
   involving the $U(1)_X$ inflaton Higgs field. 
The action in the Jordan frame relevant for the inflation scenario is given by 
\bea
 S_J= \int d^x \sqrt{g} 
 \left[ - \frac{1}{2} (M_P^2 + 2 \xi \Phi^\dagger \Phi ) {\cal R} + {\cal L}_\Phi \right], 
\eea
where ${\cal L}_\Phi$ denotes the kinetic term and the potential term (Eq.~(\ref{potential}))
  for the $U(1)_X$ field $\Phi$, and $\xi$ is a non-minimal gravitational coupling. 
The inflation trajectory is parametrized by the real component of the Higgs field, 
   $\phi=\sqrt{2} {\rm Re}[\Phi]$. 
A suitable choice of $\xi$ leads to 
  the inflationary predictions consistent with the Planck 2018 results \cite{Planck2018}. 
For example, as a benchmark, if we set $\xi=10$, which corresponds to $\lambda_X=4.0 \times 10^{-8}$, 
  we obtain the predictions for the spectral index $n_s=0.968$ and the tensor-to-scalar ratio $r=0.003$, 
  for the e-folding number $N_e=60$ (see Ref.~\cite{Okada:2011en}).

For the benchmark parameter set, we find the inflaton mass $m_\phi= \sqrt{2 \lambda_X} v_X =2.2 \times 10^6$ GeV. 
The inflaton mainly decays to a pair of SM Higgs doublets through the mixed quartic interaction in Eq.~(\ref{potential}). 
The inflaton decay width is given by 
\bea
\Gamma_\phi \simeq  \frac{\lambda_{mix}^2 \, v_X^2 }{8 \pi \, m_ \phi}, 
\label{eq:gamma2}
\eea
where we have neglected the Higgs boson mass. 
Employing this inflaton decay width, we estimate the reheating temperature by
\bea
  T_R  %=\left(\frac{90}{\pi^2 g_*}\right)^{1/4} 
  \simeq \sqrt{\Gamma_\phi M_P} 
   \simeq    10^{10} \,{\rm GeV} \, \left(\frac{\lambda_{mix}}{1.1 \times 10^{-5}}\right).      
\label{eq:TR}
\eea
Therefore, by setting, for example, $\lambda_{mix} =1.1 \times 10^{-5}$ and $Y_M^1 =0.1$, 
  we can satisfy the conditions for the successful thermal leptogenesis.

In summary, we have considered a $U(1)_X$ extended SM and proposed a light $Z^\prime$ DM by setting $x_H=-2$. 
In this case, the $U(1)_X$ gauge boson $Z^\prime$ becomes sufficiently long-lived 
   if it is lighter than the electron and thus a DM candidate in our universe.  
This scenario can explain the recently reported XENON1T excess in the electronic-recoil events 
   with a suitable choice of the model parameters, $m_{Z^\prime}=2.8$ keV and $g_X=1.8 \times 10^{-16}$, 
   compatible with the astrophysical constraints. 
Since the $U(1)_X$ gauge coupling is too small to reproduce the observed DM relic density, 
we have proposed to extend the minimal $U(1)_X$ model 
   to a two-component DM scenario by introducing 
   a Higgs-portal scalar DM, which is a SM gauge singlet but has a $U(1)_X$ charge $Q_\varphi$. 
We have found that if $Q_\varphi$ is sufficiently large, the $Z^\prime$ DM density 
   is compatible with the observed DM density by the freeze-in mechanism. 
An adequate amount of $Z^\prime$ DM particles are produced through the pair annihilation
   of the Higgs-portal DM particles in the thermal plasma.   
Our model can account not only for the XENON1T excess with the $Z^\prime$ DM, 
   it also accommodates neutrino oscillation and  implements successful inflation and letptogenesis.

%%%%%%%%%%%%%%%%%%%%%%%%%%%%%%%%%%%%%%%%%
%\section*{Acknowledgements}
{\bf Acknowledgements:} 
%%%%%%%%%%%%%%%%%%%%%%%%%%%%%%%%%%%%%%%%%
This work is supported in part by the United States Department of Energy grant DE-SC0012447 (N.~Okada) 
and DE-SC0013880 (D.~Raut and Q.~Shafi), and the M.~Hildred Blewett Fellowship of the American Physical Society, www.aps.org (S.~Okada).

%%%%%%%%%%%%%%

%%%%%%%%%%%%%%%%%%%%%%%%%%%%%%%%%%%%%%
%%%%%%%%%%%%%%%%%%%%%%%%%%%%%%%%%%%%%%

\end{document}